\documentclass[12pt,prd,aps]{revtex4}
\usepackage{amsfonts}
\usepackage{amsmath}
\usepackage{amssymb}
\usepackage{bm}

\begin{document}

\title{Characteristic initial value problems for integrable hyperbolic reductions of Einstein's equations
\footnote{This work is supported partly by the Russian Foundation for Basic Research (grants 02-01-00729 and 02-02-17372) and by the Russian Academy of Sciences (the program ``Nonlinear Dynamics and Solitons''}}
\author{G. A. Alekseev}
     \email{G.A.Alekseev@mi.ras.ru}
\affiliation{Steklov Mathematical Institute of the Russian
Academy of Sciences, \\
Gubkina str. 8, 119991, Moscow, Russia}

\begin{abstract}
A unified general approach is presented for construction of solutions of the characteristic initial value problems for various integrable hyperbolic reductions of Einstein's equations for space-times with two commuting  isometries in General Relativity and in some string theory induced gravity models. In all cases the associated linear systems of similar  structures are used and their fundamental solutions  admit an alternative representation  by two ``scattering'' matrices  of a simple analytical structure on the spectral plane. The condition of equivalence of these representations leads to the linear ``integral evolution equations''  whose scalar kernels and right hand sides are determined completely by the initial data for the fields specified on the two initial characteristics. If the initial data for the fields are given, all field components of the corresponding solution can be expressed in quadratures in terms of a unique solution of these quasi - Fredholm integral evolution  equations.
\end{abstract}

\maketitle

\section{Introduction}

The Einstein's equations in General Relativity as well as in some  string theory induced gravity models long ago were found to be integrable in a number of physically important cases. It is well known,  that if the space-time admits an Abelian  two-dimensional isometry group and all components of metric and matter fields {\it and their potentials} are dependent on some two of the four space-time coordinates only, the Einstein's equations for gravitational fields in vacuum as well as the Einstein - Maxwell field equations for electrovacuum fields have been reduced to integrable equations. The similar reductions of 
the Einstein - Maxwell - Weyl equations for gravitational, electromagnetic and massless two-component spinor  fields, the Einstein's equations for the ideal fluid with  $p=\varepsilon$, generalized Einstein's equations which arise as the string background equations and include electromagnetic, dilaton and axion fields, and some others are also integrable. 

The structures of all these integrable reductions of Einstein's equations can be described in a unified manner \cite{Alekseev:1999}. For this type of associated linear systems the space of local solutions of the field equations can be mapped by the space of monodromy data of a fundamental solution of associated linear system on the spectral plane. These space of monodromy data consists of a finite set of coordinate independent and arbitrarily chosen functions of spectral parameter. For example, any vacuum local solution is characterized by two and any electrovacuum solution -- by four functions of spectral parameter. These functions can be calculated (in principle, at least) for any given solution of the field equations, and for any choice of these functions there exists a corresponding solution of the field equations. Thus, 
similarly to the well known inverse scattering transform, 
the direct and inverse problems of such ``monodromy transform'' 
arise and each possess a unique solution. To find the monodromy data for any given solution (the direct problem), one has to find for this solution the corresponding solution of the associated linear system with spectral parameter. Solution of the inverse problem, i.e. a construction of the solution for any given monodromy data functions, is equivalent to solution of some linear singular integral equations whose scalar kernels and right hand sides are expressed completely in terms of given monodromy data functions \cite{Alekseev:1985},\cite{Alekseev:1987}. 

The described above construction of the monodromy transform 
suggests immediately some methods for construction of infinite 
hierarchies of particular exact solutions of these field 
equations \cite{Alekseev-Garcia:1996}, 
\cite{Alekseev-Griffiths:2000} and 
provides a principle opportunity for solution of the initial or boundary value problems for the reduced Einstein's equations \cite{Alekseev:1993}. Farther development of this approach was suggested in \cite{Alekseev:2001}, where for vacuum and electrovacuum fields some new, quasi - Fredholm form of linear integral equations substituting in this approach the nonlinear dynamical field equations was found. The structure of these integral equations is completely determined by the initial data for the fields on the mentioned just above two characteristics. 
Thus, this approach became well adapted for solution of the characteristic initial value problems, but its essential restriction was concerned with the conjecture of local analyticity of solutions. A straightforward generalization of this approach to the solutions with non-analytical behaviour of fields and its application to solution of the characteristic initial value problem for plane gravitational or gravitational and electromagnetic waves with distinct wavefronts colliding on the Minkowski background was presented in \cite{Alekseev-Griffiths:2001}. 

In this paper, a generalization of the monodromy transform approach for other hyperbolic integrable reductions of Einstein's equations  in General Relativity and string theory induced gravity models is presented.

\section{Integrable hyperbolic reductions of Einstein's  equations}
\label{sec:intro}

Each of the mentioned above integrable reductions of Einstein's equations possess the space-time metric of the form
\begin{equation}\label{Metric}
ds^2=g_{\mu\nu} dx^\mu dx^\nu +g_{ab}(dx^a+\omega^a{}_\mu dx^\mu)(dx^b+\omega^b{}_\nu dx^\nu)
\end{equation}
where $\mu,\nu,\ldots=1,2$ and $a,b,\ldots=3,4$ and the metric components $g_{\mu\nu}$, $g_{ab}$, $\omega^a{}_\mu$ are functions of two space-time coordinates $x^\mu$ only. The electromagnetic, Weyl massless spinor, dilaton and axion  fields, which can exist in these space-times, possess respectively the potentials 
\begin{equation}\label{Matter}
A_i(x^\mu),\quad \varphi_A(x^\mu),\quad \phi(x^\mu),\quad a_i(x^\mu)
\end{equation}
which are independent of the coordinates $x^a$. Here the indices  $i,j,k,\ldots=1,2,3,4$, while the spinor indices $A,B,\ldots=1,2$.) It seems useful to remined here that the metric functions $\omega^a{}_\mu$ should be nonzero, if a nontrivial Weyl spinor field presents in the space-time (\ref{Metric}). However,  $\omega^a{}_\mu$ should vanish, if the spinor field is absent ($\varphi_A\equiv 0$) and all the remaining fields together with their potentials, like, for example, the Kinnersly self-dual matrix potential for vacuum gravitational field, are independent of $x^a$.

In the ``hyperbolic'' case, all fields are functions of time and one of spacial coordinates. For this case it is convenient to use instead of $x^\mu$ some pair of null coordinates $(u,v)$, such that $g_{\mu\nu} dx^\mu dx^\nu=f(u,v) du dv$. Among these null coordinates, determined up to arbitrary transformations $\widetilde{u}=\widetilde{u}(u)$ and $\widetilde{v}=\widetilde{v}(v)$, we chose a pair of coordinates $(\xi,\eta)$ defined as follows.  As is well known, the Einstein's field equations for metrics (\ref{Metric}) and matter fields (\ref{Matter}) imply that the function $\alpha(u,v)$ defined as $\det \Vert g_{ab}\Vert\equiv \alpha^2$ should satisfy a simple equation $\partial_u\partial_v\alpha=0$. This allows to introduce another function $\beta(u,v)$ determined up to a real additive constant by the relations: $\partial_u\beta=\partial_u\alpha$ and  $\partial_v\beta=-\partial_v\alpha$. 
Then, we define $(\xi,\eta)$ as
$$
\xi(u)\equiv\beta(u,v)+\alpha(u,v),\qquad \eta(v)\equiv\beta(u,v)-\alpha(u,v).
$$
These coordinates are very convenient for investigation of integrability properties of the reduced Einstein's equations.

\section{Equivalent matrix problems}
The integrable hyperbolic reductions of Einstein equations mentioned above can be represented by an equivalent complex $N\times N$ - matrix problems ($N=2$ and $N=3$ for pure vacuum and electrovacuum cases respectively and the same is with presence of the Weyl spinor field, $N=4$ for a generalized gravity model with electromagnetic, axion and dilaton fields) formulated for the four unknown $N\times N$-matrix functions \cite{Alekseev:1999}
$${\bf U}(\xi,\eta),\,\, {\bf V}(\xi,\eta),\,\,{\bm
\Psi}(\xi,\eta,w),\,\, {\bf W}(\xi,\eta,w)
$$
where $w$ is a free complex (``spectral'') parameter. These  should satisfy the following three groups of conditions. 
The first group is a linear system for ${\bm \Psi}$ with case dependent structures of canonical forms of its coefficients
\begin{equation}\label{Linsys} 
\begin{array}{lcl}
\left\{\begin{array}{l}
2 i (w-\xi)\partial_\xi \bm{\Psi}={\bf U}(\xi,\eta)\bm{
\Psi}\\[1ex]
2 i (w-\eta)\partial_\eta \bm{\Psi}={\bf V} (\xi,\eta)
\bm{\Psi}\end{array}\right.
&\left.\vphantom{\begin{array}{l}
2 i (w-\xi)\partial_\xi \bm{\Psi}={\bf U}(\xi,\eta)  \bm{
\Psi}\\[1ex]
2 i (w-\eta)\partial_\eta \bm{\Psi}={\bf V} (\xi,\eta)
\bm{\Psi}\end{array}}
\hskip1ex\right\Vert\hskip1ex &
\begin{array}{l}
{\bf U}(\xi,\eta)={\cal F}(\xi,\eta){\bf U}_{(0)}{\cal F}^{-1}(\xi,\eta)\\[1ex]
{\bf V}(\xi,\eta)={\cal G}(\xi,\eta){\bf V}_{(0)}{\cal G}^{-1}(\xi,\eta)\end{array}
\end{array}
\end{equation}
where the transformation matrices ${\cal F}$ and ${\cal G}$ 
depend on the field components and their derivatives, while the expressions for ${\bf U}_{(0)}$, ${\bf V}_{(0)}$, given below, are constant or, at most, depend on the spinor  field current functions. The second group of conditions implies the existence for Eq. (\ref{Linsys}) of a Hermitian integral with case dependent constant matrix ${\bf\Omega}$:
\begin{equation}\label{Wequations} 
\begin{array}{rcl}
\left\{\begin{array}{l}
\bm{\Psi}^\dagger {\bf W} \bm{\Psi} = {\bf W}_0(w)\\[1ex]
{\bf W}_0^\dagger (w)={\bf W}_0 (w)\end{array}\right.&\left.
\vphantom{\begin{array}{l}
\bm{\Psi}^\dagger\cdot {\bf W}\cdot{\bf \Psi} = {\bf W}_0(w)\\[1ex]
{\bf W}_0^\dagger (w)={\bf W}_0 (w)\end{array}}
\hskip3ex\right\Vert\hskip3ex &
\displaystyle{\partial {\bf W}\over\partial w} = 4 i \bm{\Omega}
\end{array}
\end{equation}
where the components of ${\bf W}$ also include the components of the metric (\ref{Metric}) and the potentials of the matter fields (\ref{Matter}). The third group of conditions are the normalization conditions imposed at some chosen ``initial''  point $\xi=\xi_0$, $\eta=\eta_0$ on the values of $\bm{\Psi}$ and ${\bf W}(\xi_0,\eta_0,w)={\bf W}_0(w)$:
\begin{equation}\label{Nequations}
\bm{\Psi}(\xi_0,\eta_0,w)={\bf I},\quad {\bf W}_0(w)=4 i (w-\beta_0)\bm{\Omega}+{\bf G}_0
\end{equation}
where $\beta_0=(\xi_0+\eta_0)/2$ and ${\bf G}_0$ is also the case dependent constant matrix. 

In Eq. (\ref{Linsys}) the canonical forms  ${\bf U}_{(0)}$, ${\bf V}_{(0)}$ of the matrices ${\bf U}$ and ${\bf V}$ (up to a permutation of diagonal elements) are
$$\left.\begin{array}{l}
{\bf U}_{(0)}^{N=2}=\mbox{diag}\,(i,0)\\[1ex]
{\bf V}_{(0)}^{N=2}=\mbox{diag}\,(i,0)
\end{array}\hskip0.2ex\right\Vert\hskip0.2ex
\left.\begin{array}{l}
{\bf U}_{(0)}^{N=3}=\mbox{diag}\,(i+a(\xi),0,0)\\[1ex]
{\bf V}_{(0)}^{N=3}=\mbox{diag}\,(i+b(\eta),0,0)\end{array}
\hskip0.2ex\right\Vert\hskip0.2ex
\begin{array}{l}
{\bf U}_{(0)}^{N=4}=\mbox{diag}\,(i,i,0,0)\\
{\bf V}_{(0)}^{N=4}=\mbox{diag}\,(i,i,0,0) \end{array}
$$
where $a(\xi)$, $b(\eta)$ are the arbitrarily chosen spinor field current functions \cite{Alekseev:1983}. The constant matrices $\bm{\Omega}$ should be chosen in the form:
$$\bm{\Omega}^{N=2}=\left(\hskip-1.5ex\begin{array}{rr}
0&1\\-1&0
\end{array}\right)
\hskip1ex\left\Vert\hskip1ex
\bm{\Omega}^{N=3}=\left(\hskip-1.5ex\begin{array}{rrr}
0&1&0\\-1&0&0\\0&0&0 \end{array}\right)
\hskip1ex\right\Vert\hskip1ex
\bm{\Omega}^{N=4}=\left(\hskip-1.5ex\begin{array}{rrrr}
0&0&1&0\\0&0&0&1\\-1&0&0&0\\0&-1&0&0\end{array}\right)$$
The chosen simplest normalization of the field components at the initial point $(\xi_0,\eta_0)$ leads to the following expressions for ${\bf G}_0$ ($\alpha_0=(\xi_0-\eta_0)/2$):
$$ {\bf G}_0^{N=2}=\mbox{diag}\{4\alpha_0^2,4\},\quad {\bf
G}_0^{N=3}= \mbox{diag}\{4\alpha_0^2,4,1\},\quad {\bf
G}_0^{N=4}=\mbox{diag}\{4\alpha_0^2,4\alpha_0^2,4,4\} $$

\section{The analytical structure of $\bm{\Psi}$ on the spectral plane}
The solution of the matrix problem (\ref{Linsys}) -- (\ref{Nequations}) can be reduced to a decoupling pair of linear integral (quasi-Fredholm) equations with scalar kernels. For this, similarly to electrovacuum case \cite{Alekseev:2001}, \cite{Alekseev-Griffiths:2001}, we introduce two matrix functions $\bm{\Psi}_+(\xi,w)$ and $\bm{\Psi}_-(\eta,w)$ called there as ``in-states''
\begin{equation}\label{InStates}
\bm{\Psi}_+(\xi,w)=\bm{\Psi}(\xi,\eta_0,w),\qquad
\bm{\Psi}_-(\eta,w)=\bm{\Psi}(\xi_0,\eta,w).
\end{equation}
Each of them is a fundamental solution of the linear system of  ordinary differential equations supplied with the unit matrix as the initial conditions:
$$\left\{\begin{array}{l}
2i(w-\xi)\partial_\xi\bm{\Psi}_+={\bf U}(\xi,\eta_0)\cdot
\bm{\Psi}_+\\[1ex]
\bm{\Psi}_+(\xi_0,w)={\bf I}\end{array}\right.\qquad
\left\{\begin{array}{l}
2i(w-\eta)\partial_\eta \bm{\Psi}_-={\bf V}(\xi_0,\eta)\cdot
\bm{\Psi}_-\\[1ex]
\bm{\Psi}_-(\eta_0,w)={\bf I}\end{array}\right.$$
Hence, each of these ``in-states'' is determined completely by the values of the field components given on the characteristic $\xi=\xi_0$ or $\eta=\eta_0$. 

The components of $\bm{\Psi}$ are holomorphic on the spectral plane outside the cut consisting of two non-overlapping segments of the real axis $L_+$ and $L_-$, which join the points $w=\xi_0$ with $w=\xi$ and $w=\eta_0$ with $w=\eta$ respectively. Besides that, $\bm{\Psi}_+$ is holomorphic outside $L_+$ and $\bm{ \Psi}_-$ is holomorphic outside $L_-$ and near these cuts their  local structures are
\begin{equation}\begin{array}{lcl}\label{StructuresPlus}
L_+:&&\bm{\Psi}=\widetilde{\bm{\psi}}
_+(\xi,\eta,w)\otimes{\bf k}_+(w)+{\bf M}_+(\xi,\eta,w)\\
&&\bm{\Psi}_+=\widetilde{\bm{\psi}}_{0+}(\xi,w)\otimes{\bf k}_+(w)+{\bf M}_{0+}(\xi,w)\\[1ex]
\end{array}
\end{equation}
\begin{equation}\begin{array}{lcl}\label{StructuresMinus}
L_-:&&\bm{\Psi}=\widetilde{\bm{\psi}}_-(\xi,\eta,w)\otimes{\bf k}_-(w)+{\bf M}_-(\xi,\eta,w)\\
&&\bm{\Psi}_-=\widetilde{\bm{\psi}}_{0-}(\eta,w)\otimes{\bf k}_-(w)+{\bf M}_{0-}(\eta,w)
\end{array}
\end{equation}
Here $\widetilde{\bm{\psi}}_\pm$, $\widetilde{\bm{\psi}}_{0\pm}$ are $N$-component column-vectors and ${\bf k}_\pm(w)$ are $N$-component row-vectors
for the vacuum and electrovacuum cases (or with possible presence of the Weyl spinor field), i.e. when $N=2,3$. However for the mentioned above string gravity model with electromagnetic, axion and dilaton fields ($N=4$) the ``columns'' $\widetilde{\bm{ \psi}}_\pm$, $\widetilde{\bm{\psi}}_{0\pm}$ and ``rows'' ${\bf k}_\pm$ take the values in the Grassman manifild $G_{2,4}(C)$, i.e. they are $4\times 2$- and $2\times 4$-matrix functions respectively. For any (analytical) local solution of reduced Einstein's equations the $N\times N$-matrices ${\bf M}_\pm$, ${\bf M}_{0\pm}$ and the vectors or matrices ${\bf k}_\pm(w)$ are single-valued (holomorphic) functions of $w$ near the corresponding cuts $L_+$ or $L_-$, whereas $\widetilde{\bm{ \psi}}_\pm$, $\widetilde{\bm{\psi}}_{0\pm}$ are the only branching solutions. The coordinate independent vectors (matrices) ${\bf k}_\pm(w)$ are related with the monodromy data which characterize every solution  \cite{Alekseev:1987},\cite{Alekseev:1999}.

Father, for all cases, for any local solutions analytical at the initial point $(\xi_0,\eta_0)$, the branching character of $\widetilde{\bm{\psi}}_\pm$, $\widetilde{\bm{\psi}}_{0\pm}$ is solution independent:
$$\left.\begin{array}{lll}
L_+:\hskip1ex & \widetilde{\bm{\psi}}_+ =\lambda_+^{-1}\bm{\psi} _+,\hskip1ex
& \widetilde{\bm{\psi}}_{0+} =\lambda_+^{-1}{\bm{\psi}}_{0+}
\\[1ex]
L_-:& \widetilde{\bm{\psi}}_- =\lambda_-^{-1}{\bm{ \psi}}_-,\hskip1ex
& \widetilde{\bm{\psi}}_{0-} =\lambda_-^{-1}{\bm{\psi}}_{0-}
\end{array}\hskip1ex \right\Vert\hskip1ex \lambda_+=\sqrt{\displaystyle\frac{w-\xi}{w-\xi_0}},\hskip1ex \lambda_-=\sqrt{\displaystyle\frac{w-\eta}{w-\eta_0}}.
$$
where ${\bm{\psi}}_+$, ${\bm{\psi}}_{0+}$ and 
${\bm{\psi}}_-$, ${\bm{\psi}}_{0-}$ are holomorphic near $L_+$ and $L_-$ respectively; the functions $\lambda_+$ and $\lambda_-$ are holomprphic functions of $w$ outside $L_+$ and $L_-$ respectively and $\lambda_\pm(w=\infty)=1$. However, for the local solutions with non-analytical behaviour at the initial point $(\xi_0,\eta_0)$ as, for example, at the point of collision of waves with distinct wavefronts \cite{Alekseev-Griffiths:2001}, the character of branching of  
$\widetilde{\bm{\psi}}_+$, $\widetilde{\bm{\psi}}_{0+}$ and 
$\widetilde{\bm{\psi}}_-$, $\widetilde{\bm{\psi}}_{0-}$ can be more complicate and solution-dependent. Besides that, in the presence of the spinor field, the character of branching of ${\bm{ \Psi}}$ also changes and we have to substitute everywhere
$$\lambda_+\to\lambda_+ e^{i\sigma_+},\quad \lambda_-\to\lambda_- e^{i\sigma_-},\quad
\sigma_+=\frac12\displaystyle\int_{L_+} \displaystyle\frac {a(\zeta) d\zeta}{w-\zeta},\quad \sigma_-=\frac 12\displaystyle\int_{L_-} \displaystyle\frac {b(\zeta)d\zeta}{w-\zeta}.$$  
The defined above fragments of the algebraic structures of ${\bm{ \Psi}}$ and ${\bm{\Psi}}_\pm$ on the cuts $L_\pm$ will be used farther in our construction.

\section{The linear integral ``evolution equations''}
To construct the general solutions for the matrix problems (\ref{Linsys}) -- (\ref{Nequations}) we express $\bm{\Psi}$  in two  alternative forms in terms of ``scattering'' matrices $\bm{ \chi}_\pm$:  
\begin{equation}\label{Dressing}
\bm{\Psi}=\bm{\chi}_+(\xi,\eta,w)\cdot \bm{\Psi}_+(\xi,w),
\qquad 
\bm{\Psi}=\bm{\chi}_-(\xi,\eta,w)\cdot \bm{\Psi}_-(\eta,w).
\end{equation}
Using the methods described in \cite{Alekseev:1987}, \cite{Alekseev:2001}, we can show that $\bm{\chi}_-$ is holomorphic outside $L_+$, while $\bm{\chi}_+$ is holomorphic outside $L_-$ and they can be represented as the Cauchy integrals over these cuts. In the integrands, the jumps of the ``scattering'' matrices on $L_\pm$ can be expressed in terms of the functions defined in (\ref{StructuresPlus}), (\ref{StructuresMinus}). Then the condition of identity of the alternative representations (\ref{Dressing}) for $\bm{\Psi}$ gives rise to the basic system of quasi - Fredholm linear integral equations:
$$\left\{\begin{array}{l}
\bm{\phi}_+(\xi,\eta,\tau_+)- \displaystyle{\int_{L_-}} S_+(\xi,\tau_+,\zeta_-) 
\bm{\phi}_-(\xi,\eta,\zeta_-) \,d\zeta_- = \bm{ \phi}_{0+}(\xi,\tau_+)\\
  \bm{\phi}_-(\xi,\eta,\tau_-)-\displaystyle{\int_{L_+}}
S_-(\eta,\tau_-,\zeta_+) \bm{\phi}_+(\xi,\eta,\zeta_+) \,d\zeta_+ =\bm{\phi}_{0-}(\eta,\tau_-) \end{array}\right.$$ 
where $\tau_+,\zeta_+\in L_+$ and $\tau_-,\zeta_-\in L_-$; the  functions $\bm{\phi}_+(\tau_+)$, $\bm{\phi}_-(\tau_-)$ and their initial values $\bm{\phi}_{0+}(\tau_+)$, $\bm{\phi}_{0-}(\tau_-)$ are the jumps $[\widetilde{\bm{\psi}}_+]_{\tau_+}$, $[\widetilde{\bm{\psi}}_-]_{\tau_-}$ and
$[\widetilde{\bm{\psi}}_{0+}]_{\tau_+}$,
$[\widetilde{\bm{\psi}}_{0-}]_{\tau_-}$ respectively. The expressions for the  scalar kernels are 
$$\begin{array}{lcl} S_+(\xi,\tau_+,\zeta_-) 
=\displaystyle{\frac{\left({\bf m}_-(\xi,\zeta_-)\cdot {\bm{ \phi}}_{0+}(\xi,\tau_+)\right)}{i\pi(\zeta_--\tau_+)}},
&\qquad &{\bf m}_-={\bf k}_-(w)\cdot\bm{\Psi}_+^{-1}(\xi,w) 
\\
S_-(\eta,\tau_-,\zeta_+) =\displaystyle{\frac{\left({\bf m}_+(\eta,\zeta_+)\cdot
\bm{\phi}_{0-}(\eta,\tau_-)\right)}{
i\pi(\zeta_+-\tau_-)}},&&
{\bf m}_+={\bf k}_+(w)\cdot\bm{\Psi}_-^{-1}(\eta,w).
\end{array}
$$
Thus, the coefficients of our generalized integral ``evolution equations'' are completely determined by the characteristic initial data for the fields. In terms of their solution, all of the field components can be determined in quadratures. In particular, 
the $N\times N$-matrix functions ${\bf R}_\pm(\xi,\eta)$ defined by the expansions $\bm{\chi}_\pm={\bf I}+(1/w) {\bf R}_\pm+O(1/w^2)$ can be calculated as 
$${\bf R}_+=-\displaystyle{1\over \pi
i}\int_{L_-} \widetilde{\bm{\psi}}_- \otimes{\bf m}_-\,d\,\zeta_-,\qquad
{\bf R}_-=-\displaystyle{1\over \pi i}\int_{L_+}
\widetilde{\bm{\psi}}_+\otimes {\bf m}_+\,d\,\zeta_+.$$
In terms of these matrix functions, for solution of (\ref{Linsys}) -- (\ref{Nequations}) we obtain
$${\bf U}(\xi,\eta)={\bf U}(\xi,\eta_0)+2 i\partial_\xi{\bf R}_+,\quad
{\bf V}(\xi,\eta)={\bf V}(\xi_0,\eta)+2 i\partial_\eta{\bf R}_-$$
and therefore, the derived linear integral evolution equations  solve the characteristic initial value problems for considered integrable hyperbolic reductions of Einstein's equations.

\section*{Acknowledgments}
{\uppercase{T}his work is supported partly by the \uppercase{R}ussian \uppercase{F}oundation for \uppercase{B}asic \uppercase{R}esearch (grants 02-01-00729 and 02-02-17372) and by the \uppercase{R}ussian \uppercase{A}cademy of \uppercase{S}ciences (the program ``\uppercase{N}onlinear \uppercase{D}ynamics and \uppercase{S}olitons''})

\end{document}